\documentclass[usenatbib]{mn2e}
\usepackage{lineno}
\usepackage{natbib}
\usepackage{psfig}
\usepackage{graphicx}
\usepackage{amssymb}
\usepackage{color}

\author[Perera et al.]{B.~B.~P.~Perera$^{1}$, B.~W.~Stappers$^{1}$, P.~Weltevrede$^{1}$, A.~G.~Lyne$^{1}$, J.~M.~Rankin$^{2}$
\\ $^1$ Jodrell Bank Centre for Astrophysics, School of Physics and Astronomy, The University of Manchester, Manchester M13 9PL, UK
\\  $^2$ Physics Department, University of Vermont, Burlington, VT 05405, USA
}

\title[Spin-down and emission state changes in PSR B1859$+$07]{Correlated spin-down rates and radio emission in PSR B1859$+$07}

\begin{document}

\maketitle

\begin{abstract}
We study the spin-down changes of PSR B1859$+$07 over a period of more than 28 years of radio observation. We identify that the time derivative of the rotational frequency ($\nu$) varies quasi-periodically with a period of $\sim$350 days, switching mainly between two spin-down states. The profile shape of the pulsar is correlated with the $\dot \nu$ variation, producing two slightly different profile shapes corresponding to high- and low-$\dot \nu$ states. In addition to these two normal emission states,  we confirm the occasional flare-state of the pulsar, in which the emission appears early in spin phase compared to that of the common normal emission. The profile of the flare-state is significantly different from that of the two normal emission states. The correlation analysis further shows that the flare-state is not directly linked with the $\dot \nu$ changes. With a simple emission beam model, we estimate the emission altitude of the normal emission to be 240~km, and explain the origin of the flare-state as an emission height variation from the leading edge of the beam. We also argue that the emission of these states can be explained with a partially active beam model. In this scenario, the trailing portion of the radio beam is usually active and the normal emission is produced. The flare-state occurs when the leading edge of the beam becomes active while the trailing part is being blocked. This model estimates a fixed emission altitude of 360~km. However, the cause of the flare-state (i.e. the emission height variation, or the time-dependent activity across the radio beam) is not easily explained.
\end{abstract}

\begin{keywords}
  stars:neutron -- pulsars
\end{keywords}

\section{Introduction}

Pulsars are rapidly rotating highly magnetized neutron stars. By losing their rotational kinetic energy, pulsars spin-down gradually over time. However, the spin-down is not uniform in general, and shows irregularities. This can be clearly seen in pulsar timing residuals -- the difference between the time of arrival (TOA) of the pulse from the pulsar at the observatory and that predicted by a spin-down model. The spin irregularities are mainly due to glitch events, a sudden increase in rotational frequency \citep[see][]{els+11,ymh+13}, and low frequency timing noise. Glitches are likely common for young pulsars and their spin irregularities are due to the recovery from their glitch events, while the older pulsars show quasi-periodic variations in their timing residuals \citep{hlk10}. By using long-term observations, \citet{lhk+10} presented the spin-down rate ($\dot \nu$) changes of 17 pulsars, and explained the timing noise as being mainly due to switches between two spin-down rates over time. Some pulsars in their study show quasi-periodic $\dot \nu$ variations with long time-scales, $\geq$1~yr. Even more interestingly, they found that a few of these pulsars show $\dot \nu$-related pulse profile shape variations, suggesting that $\dot \nu$ variations are linked with some phenomenon in the magnetosphere. The two states (radio-loud `on' and radio-quiet `off' states) of the `intermittent pulsars' are also associated with different $\dot \nu$ states \citep[see][]{klo+06,llm+12,crc+12,ysl+13}, further strengthening the notion that the $\dot \nu$ changes are linked with some magnetospheric effects. All these studies reveal that $\dot \nu$ variations of pulsars are common, although the mechanism behind this is not well understood. Two mechanisms have been proposed: free precession of pulsars \citep{Jon12,alw06} and the neutron star magnetoshpere switching between two magnetospheric configurations \citep[vacuum and force-free limits, e.g.][]{lst12a}. However, the reliability and the approximations that have been made in these mechanisms are questioned \citep[see][]{lst+12,kkh+12,khk12a}.

\citet{rrw06} reported the bistable emission states of PSRs B0919$+$06 and B1859$+$07. For these pulsars, they identified the common `normal' emission state, in which the pulsar emits its radio emission most of the time, and a less common `flare' emission state, in which the emission appears slightly early in spin phase compared to that of the normal-state. They showed that these states produce two significantly different pulse profile shapes for both pulsars. The transition from the normal-state to the flare-state (and vice-versa) happens gradually over time within few pulses. \citet{psw+15} studied more than 30 years of PSR B0919$+$06 radio data and confirmed the normal- and flare-state of the pulsar, and the quasi-periodic $\dot \nu$ variation reported first in \citet{lhk+10}. 
They further found that the pulsar shows two additional emission states within the normal-state, resulting in three emission states in total. 
They noted that these two normal-states are correlated with the $\dot \nu$ variation of the pulsar, but the flare-state is not likely correlated (or the correlation is not entirely clear). 
The two normal-states are related with the high-$\dot \nu$ and low-$\dot \nu$ states of the pulsar, and the pulse profile shape difference between  these two states is very small compared to that of the flare-state and the normal-states.
\citet{psw+15} further found that the $\dot \nu$ repeating pattern is complex, each cycle consists of two switches between the two spin-down states (see Figure~3 therein). Within a cycle, $\dot \nu$ varies from the lower spin-down state to the upper spin-down state twice with different amounts of time spent in each state, resulting in a further quasi-stable secondary modulation in the $\dot \nu$ switching. 
As reported by \citet{rrw06}, PSR B1859$+$07 shows more flares than B0919$+$06 within a given time. Therefore, PSR B1859$+$07 may be a good source to study the flare-state emission and its characteristics, potentially leading to an understanding of the common features of flare-state pulsars. By comparing the radio emission of the two pulsars, it is clear that the flare-state of PSR B1859$+$07 is not stable, rather varying in both the duration ($\sim$5--30~s) and the size of the pulse phase shift ($\sim$0.01--0.04) from flare-state to flare-state \citep[see][]{rrw06}. In this work, we use the long-term radio observations of PSR B1859$+$07 to investigate its $\dot \nu$ variation, different emission states (i.e. flare-state and possible other states within the normal emission), and the correlation between these states and the $\dot \nu$ variation, similar to PSR B0919$+$06.

By using the information contained in the pulse profiles, it is possible to study the emission characteristics of pulsars. Although the pulsar emission mechanism is not well understood, we generally assume the magnetic field in pulsar magnetospheres is dipolar and the radio emission is generated by accelerating plasma along the magnetic field lines. The acceleration happens in the particle-depleted regions (so called `gaps') within the magnetosphere. The radio emission is thought to be generated within the polar cap (PC) region (i.e. the region covered by open field lines which cross the light-cylinder radius -- $R_{LC} = cP/2\pi$, where $P$ is the period of the pulsar and $c$ is the speed of light) at a low altitude close to the neutron star surface. Different magnetosphere models and techniques have been used to model the radio emission of pulsars \citep[e.g.][]{rs75,dh82,spi06,lst+12}. The geometric parameters of isolated and binary pulsars have been constrained by using polarization measurements with the rotating-vector-model \citep[see][]{rc69a,ran83,ran83a,ran90,ran93,mr11,rwj15} and by utilizing models for spin precession \citep[see][]{kra98,sta04,mks+10,bkk+08,pmk+10,plg+12,fsk+13,pkm+14}, respectively. The typical emission altitude of pulsars is about $<$10\% the light-cylinder radius \citep[see][]{gk93,kg97,hsh+12}. In this work, we use a simple analytical circular radio beam model with the inferred geometry from the polarization given in \citet{rrw06} to estimate the emission altitudes of PSR B1859$+$07 for the flare- and normal-states.

In Section~\ref{obs}, we present our observations, data processing methods, and analysis. We separate the normal- and flare-state emission and study their pulse profiles in Section~\ref{profs}. Then we analyze the spin-down rate variation and the $\dot \nu$-related pulse profile shape variation in Section~\ref{spindown}. We also study the pulse profile shapes of other states within the normal-state emission in detail. In Section~\ref{altitude}, we estimate the emission altitudes corresponding to the flare- and normal-states, and also argue the possibility of having a partially active circular beam. Finally in Section~\ref{dis}, we summarize our results.

\section{Observations}
\label{obs}

We use radio observations of PSR B1859$+$07 obtained from the Lovell Telescope at the Jodrell Bank Observatory since 1987 November. The observations were made roughly once every 20 days, resulting in about 500 epochs. The data were recorded using the `analog filter bank (AFB)' backend from 1987 November to 2009 August, and the `digital filter bank (DFB)' backend since then. 
The earlier AFB observations were carried out at multiple frequencies 400, 600, 925, and 1400~MHz (L-band), while the later AFB and all the DFB observations were made mainly at L-band. The average observation length of an epoch is about 6~min. 
The AFB and DFB data were recorded with 512 and 1024 bins across the pulse profile, written out as 1-min and 10-s sub-integrations, respectively. A sub-integration is formed by integrating all the single pulses recorded within the sub-integration length. 
We use all multiple frequency observations in the timing analysis.
Since the timescale of the flare-state is small ($\sim$6 sec), we use only the DFB data for the pulse profile analysis of different emission states. The emission of the pulsar is significantly weaker than that of PSR B0919$+$06, so that it is difficult to resolve the flare-state in the low-resolution AFB data. Unfortunately, since the DFB sub-integration length is 10~s and the flare-state timescale is comparably small, the flare-state can be mixed with the normal emission affecting the separation of sub-integrations in the construction of the pulse profiles of normal- and flare-state. However, we use both the low-resolution AFB and high-resolution DFB data in the $\dot \nu$ variation analysis to cover a long time span.

In addition to the Lovell observations, we use high S/N Arecibo Telescope observations obtained on four different days: two recent observations on MJDs 56377 and 57121, and the two data sets used in \citet{rrw06} MJDs 52739 and 53372 for comparison. These four observations were made with different durations: about 10, 20, 70, and 130 min on MJDs 52739, 53372, 56377, and 57121, respectively. The earlier two observations were made with the `Wideband Arecibo Pulsar Processor (WAPP)' and recorded four 100-MHz channels centered at 1275, 1425, 1525, and 1625~MHz. The the later two observations were carried out using the `Mock' spectrometers and recorded four 86-MHz bands centered 5~MHz lower than at same frequencies as in the earlier two observations. In addition to the Lovell observations, these high S/N observations are useful to study the normal- and flare-state pulse profiles.

\citet{psw+15} showed that the timescale of the flare-state of PSR B0919$+$06 is roughly constant. In contrast, we see in our observations that the flare-state timescale of PSR B1859$+$07 varies from flare-state to flare-state, from short flares of few pulses (i.e. $\sim$5 sec) to longer flares of few tens of pulses (i.e. $\sim$30 sec), confirming the results reported in \citet{rrw06}. The shift in spin phase of the flare emission also varies from flare-state to flare-state, up to a maximum shift of about $\sim$0.04 (i.e. about $15^\circ$ in spin longitude).

\section{Normal- and Flare-state pulse profiles}
\label{profs}

As shown in \citet{rrw06}, 
the flare-state emission can be easily identified from the normal-state emission due to its noticeable pulse phase shift.
Thus, we first study the pulse profiles of these two clearly separable emission states in this section.
As mentioned in \citet{psw+15}, the difference between the two normal-state profiles of PSR B0919$+$06 is very small and cannot be easily seperated by eye (see Figure~8 therein).
Therefore, other possible emission states within the normal emission are investigated with the $\dot \nu$ variation in Section~\ref{spindown}.

\begin{figure}
\begin{center}
\psfig{file=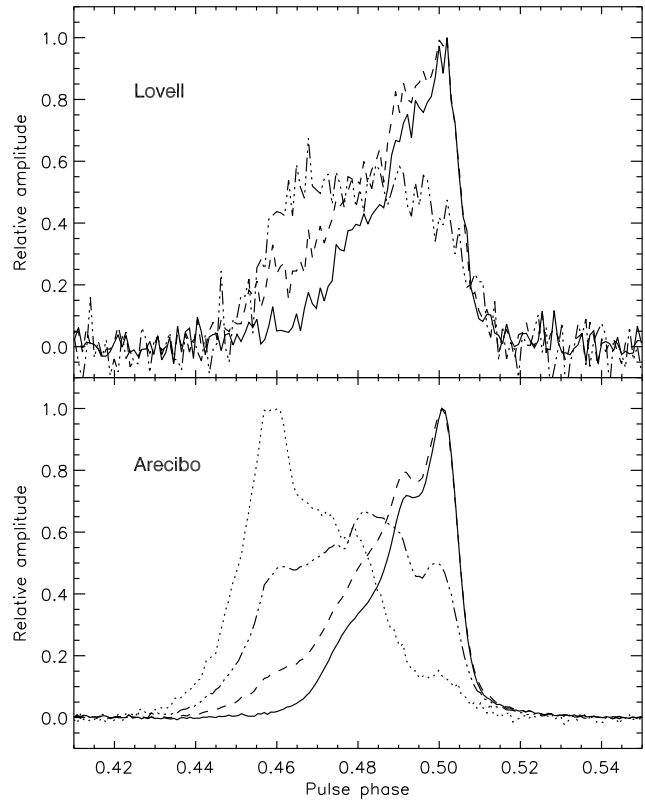,width=3.3in,angle=0}
\end{center}
\caption{
Pulse profiles of the normal-state ({\it solid}) and the flare-state ({\it dash-dot}) obtained from the Lovell DFB ({\it top}) and Arecibo ({\it bottom}) data. The {\it dashed} curve shows the total pulse profile including both states. Only 15\% of the pulse phase is shown for clarity. Note that the peak height of the normal-state profile is normalized to unity and the flare-state profile peak height is scaled according to the peak flux density ratio.
The {\it dotted} curve in the bottom panel shows the flare-state-only profile (i.e. without the transition -- see Figure~\ref{single_pulses_AO}). The peak height of this profile is normalized to unity. 
\label{profiles_DFB_AO}}
\end{figure}

We first use the DFB Lovell data to construct the flare- and normal-state pulse profiles. We separate the flare and normal emission sub-integrations of each observation by eye, and then add all these separated data to form the pulse profiles of the two states accordingly. Figure~\ref{profiles_DFB_AO} shows the pulse profiles of the two states. 
Note that since we use all non-flare emission to form the normal-state pulse profile, it may contain other possible emission states that cannot be separated by eye (see Section~\ref{spindown}).
We use all our observations to form these integrated profiles, including 716 ($\sim$2~hrs) and 225 ($\sim$38~mins) sub-integrations in the normal- and flare-states, respectively. Due to the significant difference in data lengths in the two profiles, the normal-state profile has a signal-to-noise ratio (S/N) of about two times greater than that of the flare-state profile. The peak height ratio of the normal- to flare-state profile is about 1.5, and the ratio of the power (i.e. the area under the pulse) of the two profiles is almost unity. We note that due to the relative low S/N of our DFB data, analyzing the shape variation over time is difficult.

In order to investigate and compare the pulse shapes of the profiles of the two states in detail, we use the Arecibo high S/N data obtained on 9 April 2015 (MJD 57121) which recorded single pulses. The reason to use this data is that it has the longest observation length ($\sim$2 hrs) among the four Arecibo observations. We again followed the same procedure to separate normal- and flare-states as given for the Lovell data, but use the single pulses instead of sub-integrations. Therefore, separating the two states is more clean and efficient and it is possible to minimize the mixing of the two states when constructing the corresponding pulse profiles. Figure~\ref{profiles_DFB_AO} shows the profiles of the two states. Approximately 7200 single pulses were used to form the normal-state pulse profile, resulting in $\sim$80~mins of data. In contrast, approximately 2900 single pulses were used to construct the flare-state profile, resulting in $\sim$30~mins of data. Consequently, the S/N of the normal-state profile is about 1.6 times higher than that of the flare-state profile. The peak height ratio of normal- to flare-state is about 1.5, and the power of the flare-state profile is about 1.4 times higher than that of the normal-state profile. By comparing the profiles obtained from the two telescopes, it is clear that the Arecibo data has a high S/N compared to the Lovell data, but the overall shape is similar.

The flare-state profile obtained from the Arecibo data is slightly broader, especially at the leading edge, compared to that of the Lovell data. That is because the Lovell data consists of 10~s sub-integrations, so that the flare-states with smaller timescales ($\sim$6 sec) have not been as reliably identified as flare-states during the state separation and so not included in the flare-state profile. 
In contrast, the Arecibo data are recorded with single pulses, so that no such mixing of states occurs. Therefore, the pulse profiles obtained from the Arecibo data represent actual pulse shapes of the flare- and normal-state of the pulsar. Thus, we use these profiles in Section~\ref{altitude} to estimate the emission altitudes of the pulsar.

\begin{figure}
\begin{center}
\psfig{file=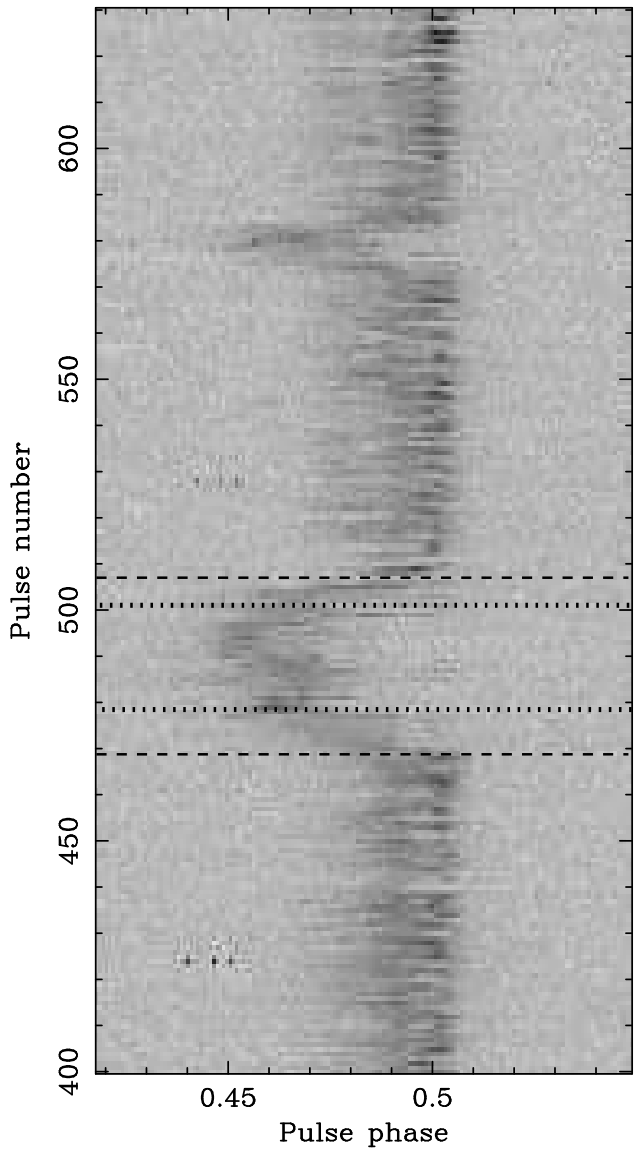,width=3.3in,angle=0}
\end{center}
\caption{
The single pulses of the pulsar within a small section ($\sim$2.5~min) of observation made on MJD 57121 using the Arecibo Telescope. Two flare-states are shown with different timescales. The longer flare-state is about 25~sec long and the shorter one is about 7~sec long. The {\it dashed} lines show the start and the end of the longer flare-state. The {\it dotted} lines show the same, but without the transition pulses. The maximum shift in pulse phase for both flares shown here are about the same, but there are flares with smaller shifts.     
\label{single_pulses_AO}}
\end{figure}

The transition between the normal- and flare-state is gradual. Figure~\ref{single_pulses_AO} shows single pulses of the pulsar during a section of the observation on MJD 57121, including two flare-states of different timescales. The longer flare-state is centred around pulse number 490. The transition starts around pulse number 470 and gradually the emission moves to early pulse phases and reaches a roughly stable state around pulse number $\sim$480. Then the emission remains relatively stable for about 20 pulses and the flare-state ends with a gradual transition back to the normal-state. The maximum phase shift is $\sim$$0.04$ in pulse phase (i.e. about $15^\circ$ in pulse longitude) and the timescale of the flare, including the transition, is about 40 single pulses (i.e. about 25~sec). The shorter flare occurs around pulse number 580 and its timescale is about 10 pulses, so that the structure of the transition is less clear. We also notice that some flare-states show small shifts in pulse phase about 0.02. Thus, the flare-state itself and its transition are complicated and somewhat similar to weak and strong flares found by \citet{psw+15} for PSR B0919$+$06. However, the occurrence of flare-states is more frequent for B1859$+$07 compared to those of B0919$+$06.

\begin{figure}
\begin{center}
\psfig{file=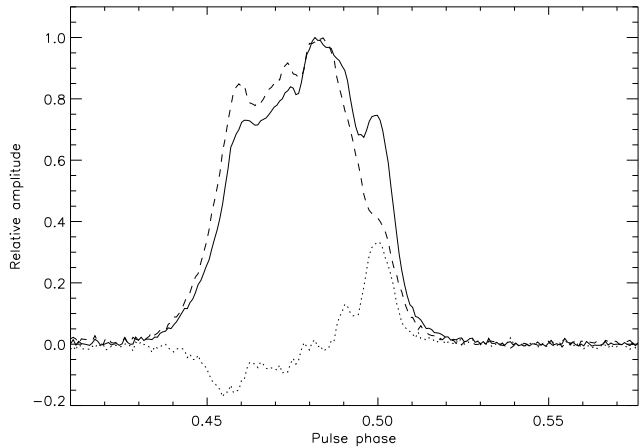,width=3.3in,angle=0}
\end{center}
\caption{
The flare-state integrated pulse profiles obtained with the Arecibo observations made on MJDs 56377 ({\it dashed}) and 57121 ({\it solid}). The alignment of the two profiles is based on the cross-correlation of the normal-state pulse profiles of the two observations. The peak heights of both profiles are normalized to unity. The difference between the profiles is shown by the {\it dotted} line.
\label{flare_profs}}
\end{figure}

Moreover, we define the flare-state starting from the beginning of the transition from the normal-state until the end of the transition back to the normal-state. Therefore, the flare-state pulse profile trailing edge ends at the same pulse phase as that of the normal-state profile. This is common for both Lovell and Arecibo data (see Figure~\ref{profiles_DFB_AO}). 
As seen in Figure~\ref{single_pulses_AO}, the timescale and the shift in pulse phase from the normal emission varies from flare-state to flare-state. Therefore, the flare-state integrated pulse profile varies from observation to observation. To show this, we constructed the flare-state integrated pulse profile from the Arecibo observations made on MJDs 56377 and 57121 (see Figure~\ref{flare_profs}), including approximately 2868 and 1516 pulses in each observation, respectively. It is clear that the profile shapes are different. The shape of the profile depends on how many flare-states there are in the observation, the emission features within the flare-state, and what criteria is used in separating the flare and normal emission states.

We further note that the flare-state-only profile, constructed without transition pulses, is somewhat different (see the bottom panel of Figure~\ref{profiles_DFB_AO}). In order to form this profile, we select flare-state-only pulses (i.e. no pulses from the transition -- see dotted lines in Figure~\ref{single_pulses_AO}) by eye and add them together. As shown in Figure~\ref{profiles_DFB_AO}, the slope of the leading edge of this profile is steeper than that of the trailing edge and the overall shape is like a mirror image of the normal-state profile. Due to low resolution and 10~s sub-integrations, it is difficult to separate flare-state-only pulses in the Lovell observations, so that we use only the Arecibo observation to construct this profile.

\section{Spin-down rate and $\dot nu$-related pulse profile shape variation}
\label{spindown}

\begin{figure*}
\begin{center}
\psfig{file=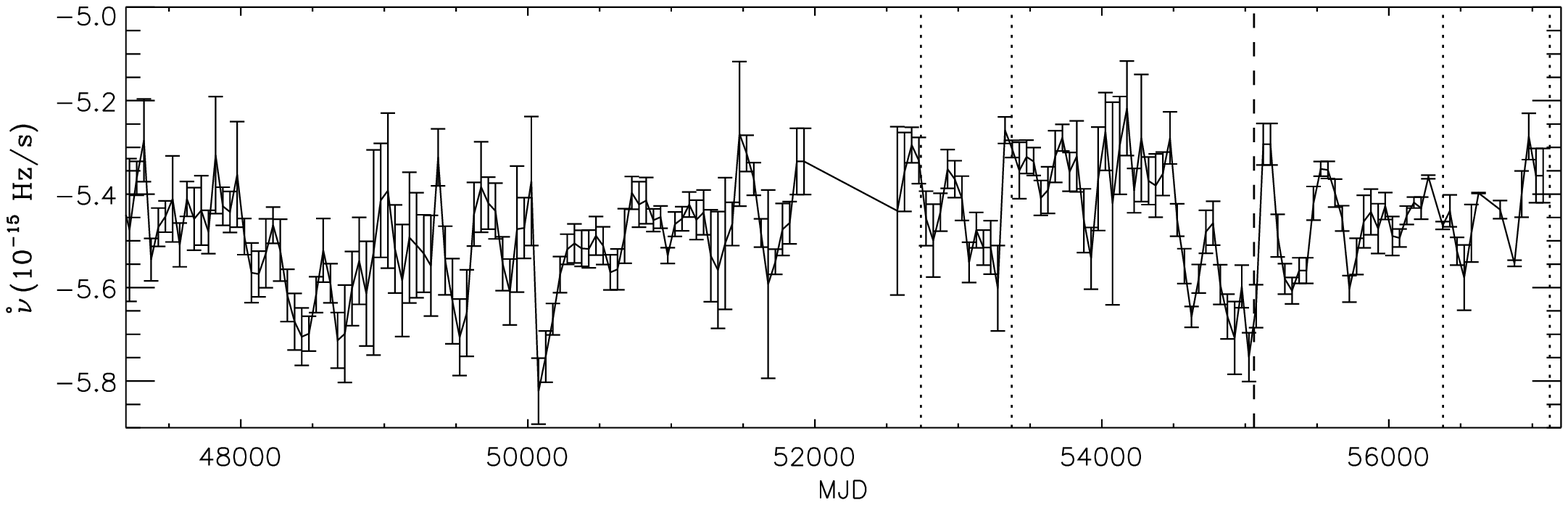,width=7in,angle=0}
\end{center}
\caption{
The $\dot \nu$ variation over time of PSR B1859$+$07 obtained from the Lovell Telescope data. The AFB data were used until MJD 55060 (marked by a {\it dashed} line) and DFB since then. We use 250-day sections with 50-d strides in the timing to generate these data points. Note that the uncertainties of $\dot \nu$ of the AFB data are relatively larger compared to that of the DFB data due to low resolution. The gap around MJD 52000 is due to an extended maintenance period of the Lovell Telescope. The {\it dotted} lines show the epochs of observations that were made with the Arecibo Telescope.  
\label{nu_dot}}
\end{figure*}

\begin{figure}
\begin{center}
\psfig{file=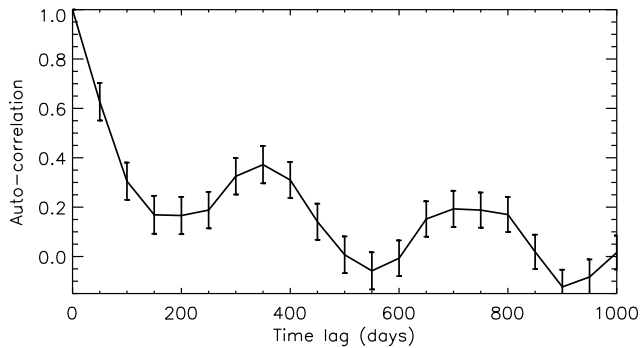,width=3.3in,angle=0}
\end{center}
\caption{
The auto-correlation function of the $\dot \nu$ variation given in Figure~\ref{nu_dot}. The periodicity of the $\dot \nu$ variation is estimated to be $\sim$350 days. 
\label{auto}}
\end{figure}

We use both AFB and DFB Lovell data to study the long-term spin-down rate variation of the pulsar. We follow the same procedure as \citet{psw+15} for PSR B0919$+$06. The $\dot \nu$ values were obtained by using timing solutions for subsequent partially overlapping sections of data (hereafter `stride fit'). After testing several trials, we find 250-day sections with 50-day strides provide the best representation of the $\dot \nu$ variation over the entire observation length of more than 28 years. This results in a quasi-periodic variation in $\dot \nu$ (see Figure~\ref{nu_dot}). By auto-correlating the $\dot \nu$ time sequence, we find that the periodicity of the $\dot \nu$ variation is $\sim$350 days (Figure~\ref{auto} shows the correlation function).
To estimate the uncertainties of the coefficients, we randomize the order of the $\dot \nu$ rates 1000 times and obtain the auto-correlation for each trial. We then calculate the standard deviation of each time lag to get the uncertainty.
As seen in Figure~\ref{nu_dot}, the high time resolution of the DFB data compared to early AFB data results in smaller uncertainties in $\dot \nu$ after 2009 August (MJD 55060). These DFB data suggests two spin-down states of the pulsar with quasi-periodic switching between them. This two-state spin-down variation is similar to the results of \citet{lhk+10} and \citet{psw+15}. Since we use partially overlapping sections of data, we note that the $\dot \nu$ values and their variations in the figure represent a smoothed average.

\begin{figure}
\begin{center}
\psfig{file=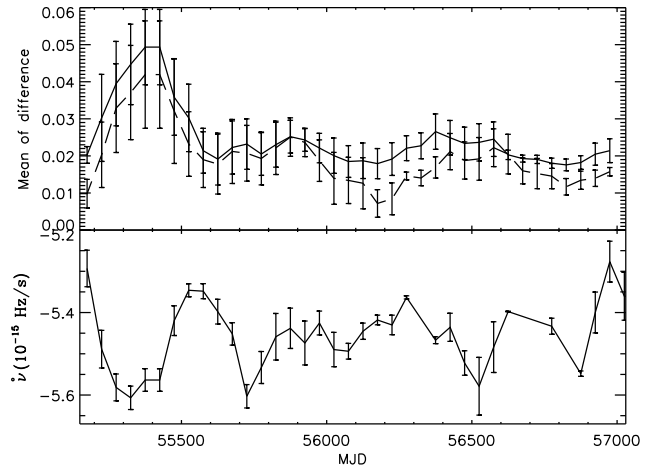,width=3.3in,angle=0}
\end{center}
\caption{
The averaged shape parameter variation over time for flare-included ({\it solid}) and flare-subtracted ({\it dashed}) profiles obtained from the DFB Lovell data. We show the $\dot \nu$ variation in the bottom panel for comparison.
\label{shape_par}}
\end{figure}

We then investigate the profile variation over time using the DFB Lovell data since 2009 August. To do this, we obtain the integrated pulse profile for each observation epoch and then resample to 256 bins (the data originally has 1024 bins across the profile) to increase the S/N per bin.  As described in other studies \citep{lhk+10,psw+15}, we first attempt to fit several Gaussians to the observed pulse profile, and then use the noise free Gaussian composite profile to obtain the pulse shape parameters. However, this approach was not successful and Gaussian fits were not good enough due to the low S/N of the observed pulse profiles. Thus, we use a profile subtracting method to estimate the profile shape variations over time. We first obtain the total normal-state integrated profile by adding only the normal emission sub-integrations from all observations, and then subtract it from each individual pulse profile to get the difference in profile shapes. We normalize the peaks of the two profiles to unity before subtracting each other. Then we calculate the mean of the difference-profile and use it as the pulse shape parameter of the given observation epoch. In order to test the effect of the flare-state on pulse profile shape, we follow this method for both flare-subtracted individual profiles (i.e. the profiles constructed by only using the normal emission sub-integrations) and flare-included individual profiles (i.e. the profiles constructed by including both emission). Figure~\ref{shape_par} shows the average shape parameter variation after applying the stride fitting for both flare-included and flare-subtracted profiles. Since we use the total flare-subtracted pulse profile to get profile differences for each epoch, the values of mean-differences are positive. The uncertainties of the averaged shape parameters of the figure are calculated from the standard deviation of the individual shape parameters within each 250-day section.

\begin{figure}
\begin{center}
\psfig{file=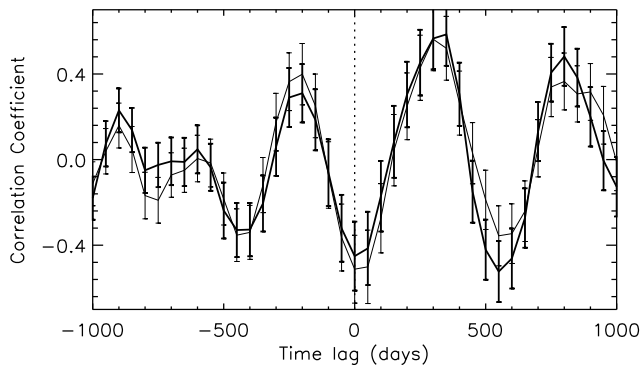,width=3.3in,angle=0}
\end{center}
\caption{
The correlation coefficient between the $\dot \nu$ variation and the shape parameter variation obtained including the flare-states ({\it thick solid}) and excluding the flare-state ({\it thin solid}).
\label{cc}}
\end{figure}

We then investigate the link between the $\dot \nu$ variation and the pulse profile shape variation by cross-correlating the $\dot \nu$ variation curve and the shape parameter variation curve to obtain the correlation coefficients (see Figure~\ref{cc}). To estimate the uncertainties of the coefficients, we randomize the order of the shape parameters 1000 times and cross-correlate with the $\dot \nu$ curve each time. Then we calculate the standard deviation of the coefficient of each time lag to get the uncertainty. We find the correlation coefficient at zero time lag for the flare-included and flare-subtracted profiles are $-0.45\pm0.16$ and $-0.51\pm0.16$, resulting in about 2.8$\sigma$ and 3.2$\sigma$ correlation, respectively. According to these results, it is clear that the flare-state is not fully associated with the $\dot \nu$ variation. 
Similar to the results found in \citet{psw+15} on PSR B0919$+$06, if the flare-state is connected with the spin-down states, then we should not see a correlation between $\dot \nu$ variation and the flare-subtracted shape parameters.

\begin{figure*}
\begin{center}
\psfig{file=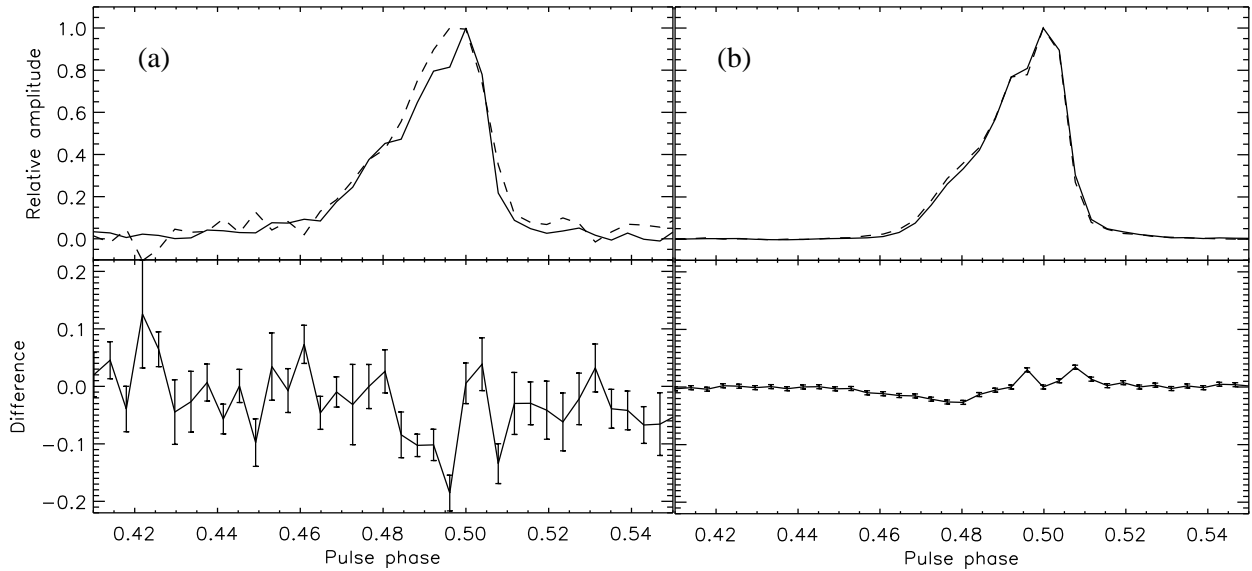,width=6.5in,angle=0}
\end{center}
\caption{
(a) The flare-subtracted pulse profiles corresponding to high ({\it solid}) and low ({\it dashed}) spin-down states obtained from the DFB Lovell observations. (b) The flare-subtracted pulse profiles obtained from the Arecibo observations made on MJD 56377 ({\it solid}) and MJD 57121 ({\it dashed}) that correspond to two different $\dot \nu$ rates close to the high spin-down state (see Figure~\ref{nu_dot}). Note that the peak of the profiles are normalized to unity. The {\it bottom} panels show the difference between the two profiles shown in the {\it top} panels. Both DFB Lovell and Arecibo profiles have 256 bins across the entire pulse phase and a fraction of the pulse phase is shown for clarity.  
\label{modes}}
\end{figure*}

Regardless of the flare-state, Figure~\ref{cc} shows a correlation between the $\dot \nu$ states and the pulse profile shape. 
This reveals that the pulsar shows more emission states corresponding to the spin-down during the normal emission, similar to PSR B0919$+$06 \citep[see][]{psw+15}.
To study the profile shape difference between the high spin-down state and the low spin-down state, we use a cut-off spin-down rate of $\dot \nu_{\rm cut} = -5.5\times10^{-15}$~Hz$/$s, which is the long-term average of the measured $\dot \nu$. When $\dot \nu > \dot \nu_{\rm cut}$, we consider the pulsar is at the high spin-down state. By integrating all high spin-down state profiles, we obtain the total high spin-down state pulse profile. Similarly, we obtain the total low spin-down state pulse profile by adding all the individual profiles when $\dot \nu < \dot \nu_{\rm cut}$. We only use flare-subtracted profiles in each case to form the two integrated profiles accordingly. Figure~\ref{modes}(a) shows the two pulse profiles corresponding to the high and low spin-down states and their difference. Even though the profiles have low S/N, it is seen that there is a difference in the pulse profiles, around pulse phase $\sim$$0.49$. This confirms the $\dot \nu$-related pulse profile variation of the pulsar. The uncertainty of the difference is obtained by randomizing the two sets of individual profiles corresponding to high and low spin-down states and then calculating the standard deviation of the intensity variation in each pulse phase bin based on 1000 trials.

To study further the $\dot \nu$-related pulse profile shapes, we try to use the Arecibo observations. However, Figure~\ref{nu_dot} shows all our Arecibo observations were made when the pulsar was within (or close to) the high-$\dot \nu$ state (see dotted lines). Thus, we cannot use the Arecibo data to study the profile shape difference between the high and low spin-down states. Nevertheless, we select the two most recent Arecibo observations (MJDs 56377 and $57121$), which have the longest observation lengths, to test for the presence of any profile shape difference. Figure~\ref{modes}(b) shows the pulse profiles of the two epochs and the difference between them. The uncertainty of the difference is calculated from the off-pulse standard deviation of the two profiles. The difference of the two profiles obtained from Arecibo observations is very small compared to that obtained from the DFB Lovell observations (see the bottom panels in Figure~\ref{modes}). 
This suggests that the two Arecibo observations are indeed in the high-$\dot \nu$ state. For comparison, we measure the profile difference between other combinations of Arecibo observations and find similar results.

\section{Possible explanations for the flare-state emission}
\label{altitude}

As mentioned in \citet{rrw06}, an emission altitude variation between states and a shift of the partially active emission region across the PC are possible explanations for the apparent phase shift of the flare-state of the pulsar with respect to the normal emission. 
In this section, we further explore these two scenarios with our new data. We study the necessary emission altitude variation in the first scenario, assuming the entire PC region is active. The observed pulse profile shape of a pulsar basically depends on the emission structure of the radio beam and the orientation of the pulsar with respect to our line-of-sight. 
The latter can be inferred from radio polarization measurements and combined with the pulse widths obtained from the observed pulse profile.
Polarization measurements of PSR B1859$+$$07$ shown in \citet{rrw06} reveal that both normal and flare-states have a similar behavior in the polarization-angle (PA) variation across the profile. By fitting the rotating-vector-model \citep{rc69a} to the PA sweep, they infer that $\alpha = 31^\circ$ and $\zeta=35.8^\circ$, where $\alpha$ is the magnetic axis inclination angle and $\zeta$ is the view angle of the line-of-sight with respect to the spin axis of the pulsar. For a given pulse width $w$, $\alpha$, and $\zeta$, we can calculate the half opening angle $\rho$ of a circular beam by $\cos\rho = \cos\alpha \cos\zeta + \sin\alpha \sin\zeta\cos(w/2)$ for a profile centered at the magnetic axis \citep[see][]{ggr84}. Then using the small angle approximation, we can express the emission height $h = 2cP\rho^2/9\pi$ \citep[e.g.][]{gg01,rwj15}.

We use the high S/N Arecibo profile shown in Figure~\ref{profiles_DFB_AO} to measure the pulse width of the normal-state. 
Since the profile difference between high- and low-$\dot \nu$ states is very small, we consider these states together as the normal state.
Assuming that the magnetic axis (or the fiducial point) is centered at the normal-state profile, we measure the 5\% pulse width to be $w = 21.5^\circ$. By using the above expressions, we then calculate the half opening angles of the beam to be $\rho=7.6^\circ$, resulting in a symmetic beam with an emission altitude of 24~R$_{NS}$ (the radius of the neutron star is assumed to be $\rm R_{ NS} = 10$~km).

As we noticed in Figure~\ref{profiles_DFB_AO}, the leading edge of the flare-state profile appears early in spin phase, while the trailing edge is at the same phase as that of the normal-state. Therefore, to produce the early phase of the leading edge of the flare-state, the emission altitude from the leading edge of the beam should be large compared to that of the trailing edge, resulting in an asymmetric beam. A simple overall increase of the emission height would result in emission arriving early at the leading edge and late at the trailing edge, which is not observed. Therefore, the emission should be generated higher up at the leading edge of the beam, while it stays the same at the trailing edge. By using the same fiducial point (i.e. the centre of the normal-state profile), we estimate the emission altitude corresponding to the leading edge of this beam to be 50~R$_{NS}$. In other words, the emission height at the leading edge of the beam in the flare-state should increase with about 26~R$_{NS}$ to produce the flare-state emission of the pulsar, while that of the trailing edge should be twice as low. The mechanism responsible for this asymmetry is not clear.

The second scenario is to invoke the partially active beam model \citep[see][]{lm88,mks+10,mr11} to explain the flare-state emission. 
In this picture the asymmetric pulse profile of the normal-state (i.e. shallow leading edge with a steep trailing edge -- see Figure~\ref{profiles_DFB_AO}) is due to a partially active beam, where its leading portion is being blocked (or not able to produce pulsed radio emission due to a failure of coherence). This is similar to the partial cone interpretation given in \cite{lm88}. During the transition from the normal-state to the flare-state, the leading portion of the beam gradually becomes active, and finally the trailing portion of the beam becomes non-active when the flare-state is reached. The asymmetric flare-state-only pulse profile (see the dotted curve in Figure~\ref{profiles_DFB_AO} -- almost a mirror image of the normal-state profile) suggests that this scenario is a good explanation for the flare-state emission. In other words, only a portion of the beam is active at a given time and the active region moves across the beam to produce the flare-state emission. If we believe this mechanism, the integrated flare-state profile including transition pulses (i.e. the dot-dash curve in Figure~\ref{profiles_DFB_AO}) is formed when the full PC is active.
Therefore, the center of the integrated flare-state profile can be considered as the center of the radio beam. Then the half opening angle of the radio beam is $\rho=9.3^\circ$ and the corresponding emission altitude is 36~R$_{NS}$ for both states. However, the mechanism and reason for moving the active portion across the beam is not easily understood, hence it is unclear which scenario is more applicable. 
Also, any intermediate solution is possible and the starting assumption that the full PC is active at some point during the flare- or normal-state might be invalid.
If only a limited area of the full PC region is involved, then the above estimates are lower limits of the emission altitudes of the pulsar.

\section{Summary}
\label{dis}

By studying more than 28 years of observations, we find the $\dot \nu$ of PSR B1859$+$07 varies quasi-periodically over time with a period of about 350~days. Relatively high S/N DFB Lovell observations show that the pulsar is likely switching between two $\dot \nu$ states (see Figure~\ref{nu_dot}), consistent with most $\dot \nu$ varying pulsars \citep{lhk+10}. 
By studying pulse profile shape variation within the normal-state emission of the pulsar, we find that the shape variation is correlated with the $\dot \nu$ variation (see Figure~\ref{shape_par} and \ref{cc}), producing two different pulse profile shapes corresponding to high- and low-$\dot \nu$ states. 
The pulse profile shape difference between these two normal emission states is very small compared to that of a typical state switching pulsar \citep[see][]{ran86,wmj07,lhk+10}. This small profile shape difference is similar to the results of the two $\dot \nu$-related emission states of PSRs B0919$+$06 \citep{psw+15} and B1540$-$06 \citep{lhk+10}. 
The correlation analysis further shows that the flare-state emission is not directly linked with the $\dot \nu$ variation.

We estimate the emission altitude corresponding to the normal-state to be $\sim$24~R$_{NS}$ by assuming a fully active PC beam model. The flare-state is explained with this model due to an emission altitude variation at the leading edge of the beam, while keeping the trailing edge emission fixed at the normal-state emission altitude (resulting in an asymmetric beam). This suggests a leading edge emission height of $\sim$50~R$_{NS}$ during the flare-state. Alternatively, we use a partially active beam scenario to explain the emission states of the pulsar. In this model, the active region of the beam is usually located at the trailing edge of the beam and produces the normal-state emission. Sometimes the active region moves across the beam towards the leading edge with time and produces the flare-state emission. With asymmetric shapes of normal-state and flare-state-only pulse profiles (see Figure~\ref{profiles_DFB_AO}), this partially active beam model provides a better explanation for the radio emission of this pulsar.

As reported in \citet{rrw06}, PSRs B1859$+$07 and B0919$+$06 show flare-state emission in addition to the normal regular radio emission. As shown in \citet{psw+15} and in this work, the timing and profile shape behavior of both pulsars show similar features such as quasi-periodic long term $\dot \nu$ variation over time and $\dot \nu$-related pulse profile shape variation.
Therefore, it is likely that the flare-state exhibiting pulsars have similar emission characteristics, but we need a larger pulsar sample with long observation lengths and better data to claim this conclusion. Additionally, the single pulse data are required to separate different emission states and form corresponding pulse profiles accurately. The results in this work provide useful information for future observations and data analysis of flare-state pulsars.

\noindent

\bigskip

\section*{Acknowledgments}
Pulsar research at the Jodrell Bank Centre for Astrophysics and the observations using the Lovell Telescope is supported by a consolidated grant from the STFC in the UK.

\bibliography{psrrefs,modrefs,journals,0737Ack}

\begin{thebibliography}{}

\bibitem[\protect\citeauthoryear{{Akg{\"u}n}, {Link} \&
  {Wasserman}}{{Akg{\"u}n} et~al.}{2006}]{alw06}
{Akg{\"u}n} T.,  {Link} B.,    {Wasserman} I.,  2006, MNRAS, 365, 653


\bibitem[\protect\citeauthoryear{{Breton}, {Kaspi}, {Kramer}, {McLaughlin},
  {Lyutikov}, {Ransom}, {Stairs}, {Ferdman}, {Camilo} \& {Possenti}}{{Breton}
  et~al.}{2008}]{bkk+08}
{Breton} R.~P.,  {Kaspi} V.~M.,  {Kramer} M.,  {McLaughlin} M.~A.,  {Lyutikov}
  M.,  {Ransom} S.~M.,  {Stairs} I.~H.,  {Ferdman} R.~D.,  {Camilo} F.,
  {Possenti} A.,  2008, Science, 321, 104

\bibitem[\protect\citeauthoryear{{Camilo}, {Ransom}, {Chatterjee}, {Johnston}
  \& {Demorest}}{{Camilo} et~al.}{2012}]{crc+12}
{Camilo} F.,  {Ransom} S.~M.,  {Chatterjee} S.,  {Johnston} S.,    {Demorest}
  P.,  2012, ApJ, 746, 63

\bibitem[\protect\citeauthoryear{Daugherty \& Harding}{Daugherty \&
  Harding}{1982}]{dh82}
Daugherty J.~K.,  Harding A.~K.,  1982, ApJ, 252, 337

\bibitem[\protect\citeauthoryear{{Espinoza}, {Lyne}, {Stappers} \&
  {Kramer}}{{Espinoza} et~al.}{2011}]{els+11}
{Espinoza} C.~M.,  {Lyne} A.~G.,  {Stappers} B.~W.,    {Kramer} M.,  2011,
  MNRAS, 414, 1679

\bibitem[\protect\citeauthoryear{{Ferdman}, {Stairs}, {Kramer}, {Breton},
  {McLaughlin}, {Freire}, {Possenti}, {Stappers}, {Kaspi}, {Manchester} \&
  {Lyne}}{{Ferdman} et~al.}{2013}]{fsk+13}
{Ferdman} R.~D.,  {Stairs} I.~H.,  {Kramer} M.,  {Breton} R.~P.,  {McLaughlin}
  M.~A.,  {Freire} P.~C.~C.,  {Possenti} A.,  {Stappers} B.~W.,  {Kaspi} V.~M.,
   {Manchester} R.~N.,    {Lyne} A.~G.,  2013, ApJ, 767, 85

\bibitem[\protect\citeauthoryear{{Gangadhara} \& {Gupta}}{{Gangadhara} \&
  {Gupta}}{2001}]{gg01}
{Gangadhara} R.~T.,  {Gupta} Y.,  2001, ApJ, 555, 31


\bibitem[\protect\citeauthoryear{{Gil}, {Gronkowski} \& {Rudnicki}}{{Gil}
  et~al.}{1984}]{ggr84}
{Gil} J.,  {Gronkowski} P.,    {Rudnicki} W.,  1984, A\&A, 132, 312



\bibitem[\protect\citeauthoryear{Gil \& Kijak}{Gil \& Kijak}{1993}]{gk93}
Gil J.,  Kijak K.,  1993, A\&A, 273, 563


\bibitem[\protect\citeauthoryear{{Hassall}, {Stappers}, {Hessels}, {Kramer},
  {Alexov}, {Anderson}, {Coenen}, {Karastergiou}, {Keane}, {Kondratiev},
  {Lazaridis}, {van Leeuwen} \& {Noutsos}}{{Hassall} et~al.}{2012}]{hsh+12}
{Hassall} T.~E.,  {Stappers} B.~W.,  {Hessels} J.~W.~T.,  {Kramer} M.,
  {Alexov} A.,  {Anderson} K.,  {Coenen} T.,  {Karastergiou} A.,  {Keane}
  E.~F.,  {Kondratiev} V.~I.,  {Lazaridis} K.,  {van Leeuwen} J.,    {Noutsos}
  A.,  2012, A\&A, 543, A66

\bibitem[\protect\citeauthoryear{{Hobbs}, {Lyne} \& {Kramer}}{{Hobbs}
  et~al.}{2010}]{hlk10}
{Hobbs} G.,  {Lyne} A.~G.,    {Kramer} M.,  2010, MNRAS, 402, 1027

\bibitem[\protect\citeauthoryear{{Jones}}{{Jones}}{2012}]{Jon12}
{Jones} D.~I.,  2012, MNRAS, 420, 2325

\bibitem[\protect\citeauthoryear{{Kalapotharakos}, {Kazanas}, {Harding} \&
  {Contopoulos}}{{Kalapotharakos} et~al.}{2012}]{kkh+12}
{Kalapotharakos} C.,  {Kazanas} D.,  {Harding} A.,    {Contopoulos} I.,  2012,
  ApJ, 749, 2

\bibitem[\protect\citeauthoryear{{Kalapotharakos}, {Harding}, {Kazanas} \&
  {Contopoulos}}{{Kalapotharakos} et~al.}{2012a}]{khk12a}
{Kalapotharakos} C.,  {Harding} A.~K.,  {Kazanas} D.,    {Contopoulos} I.,
  2012a, ApJL, 754, L1



\bibitem[\protect\citeauthoryear{{Kijak} \& {Gil}}{{Kijak} \&
  {Gil}}{1997}]{kg97}
{Kijak} J.,  {Gil} J.,  1997, MNRAS, 288, 631



\bibitem[\protect\citeauthoryear{{Kramer}}{{Kramer}}{1998}]{kra98}
{Kramer} M.,  1998, ApJ, 509, 856

\bibitem[\protect\citeauthoryear{{Kramer}, {Lyne}, {O'Brien}, {Jordan} \&
  {Lorimer}}{{Kramer} et~al.}{2006}]{klo+06}
{Kramer} M.,  {Lyne} A.~G.,  {O'Brien} J.~T.,  {Jordan} C.~A.,    {Lorimer}
  D.~R.,  2006, Science, 312, 549

\bibitem[\protect\citeauthoryear{{Li}, {Spitkovsky} \& {Tchekhovskoy}}{{Li}
  et~al.}{2012a}]{lst12a}
{Li} J.,  {Spitkovsky} A.,    {Tchekhovskoy} A.,  2012a, ApJL, 746, L24

\bibitem[\protect\citeauthoryear{{Li}, {Spitkovsky} \& {Tchekhovskoy}}{{Li}
  et~al.}{2012b}]{lst+12}
{Li} J.,  {Spitkovsky} A.,    {Tchekhovskoy} A.,  2012b, ApJ, 746, 60

\bibitem[\protect\citeauthoryear{{Lorimer}, {Lyne}, {McLaughlin}, {Kramer},
  {Pavlov} \& {Chang}}{{Lorimer} et~al.}{2012}]{llm+12}
{Lorimer} D.~R.,  {Lyne} A.~G.,  {McLaughlin} M.~A.,  {Kramer} M.,  {Pavlov}
  G.~G.,    {Chang} C.,  2012, ApJ, 758, 141

\bibitem[\protect\citeauthoryear{{Lyne}, {Hobbs}, {Kramer}, {Stairs} \&
  {Stappers}}{{Lyne} et~al.}{2010}]{lhk+10}
{Lyne} A.,  {Hobbs} G.,  {Kramer} M.,  {Stairs} I.,    {Stappers} B.,  2010,
  Science, 329, 408

\bibitem[\protect\citeauthoryear{Lyne \& Manchester}{Lyne \&
  Manchester}{1988}]{lm88}
Lyne A.~G.,  Manchester R.~N.,  1988, MNRAS, 234, 477

\bibitem[\protect\citeauthoryear{{Manchester}, {Kramer}, {Stairs}, {Burgay},
  {Camilo}, {Hobbs}, {Lorimer}, {Lyne}, {McLaughlin}, {McPhee}, {Possenti},
  {Reynolds} \& {van Straten}}{{Manchester} et~al.}{2010}]{mks+10}
{Manchester} R.~N.,  {Kramer} M.,  {Stairs} I.~H.,  {Burgay} M.,  {Camilo} F.,
  {Hobbs} G.~B.,  {Lorimer} D.~R.,  {Lyne} A.~G.,  {McLaughlin} M.~A.,
  {McPhee} C.~A.,  {Possenti} A.,  {Reynolds} J.~E.,    {van Straten} W.,
  2010, APJ, 710, 1694

\bibitem[\protect\citeauthoryear{{Mitra} \& {Rankin}}{{Mitra} \&
  {Rankin}}{2011}]{mr11}
{Mitra} D.,  {Rankin} J.~M.,  2011, ApJ, 727, 92

\bibitem[\protect\citeauthoryear{{Perera}, {Kim}, {McLaughlin}, {Ferdman},
  {Kramer}, {Stairs}, {Freire} \& {Possenti}}{{Perera} et~al.}{2014}]{pkm+14}
{Perera} B.~B.~P.,  {Kim} C.,  {McLaughlin} M.~A.,  {Ferdman} R.~D.,  {Kramer}
  M.,  {Stairs} I.~H.,  {Freire} P.~C.~C.,    {Possenti} A.,  2014, ApJ, 787,
  51

\bibitem[\protect\citeauthoryear{{Perera}, {Lomiashvili}, {Gourgouliatos},
  {McLaughlin} \& {Lyutikov}}{{Perera} et~al.}{2012}]{plg+12}
{Perera} B.~B.~P.,  {Lomiashvili} D.,  {Gourgouliatos} K.~N.,  {McLaughlin}
  M.~A.,    {Lyutikov} M.,  2012, ApJ, 750, 130

\bibitem[\protect\citeauthoryear{{Perera}, {McLaughlin}, {Kramer}, {Stairs},
  {Ferdman}, {Freire}, {Possenti}, {Breton}, {Manchester}, {Burgay}, {Lyne} \&
  {Camilo}}{{Perera} et~al.}{2010}]{pmk+10}
{Perera} B.~B.~P.,  {McLaughlin} M.~A.,  {Kramer} M.,  {Stairs} I.~H.,
  {Ferdman} R.~D.,  {Freire} P.~C.~C.,  {Possenti} A.,  {Breton} R.~P.,
  {Manchester} R.~N.,  {Burgay} M.,  {Lyne} A.~G.,    {Camilo} F.,  2010, ApJ,
  721, 1193

\bibitem[\protect\citeauthoryear{{Perera}, {Stappers}, {Weltevrede}, {Lyne} \&
  {Bassa}}{{Perera} et~al.}{2015}]{psw+15}
{Perera} B.~B.~P.,  {Stappers} B.~W.,  {Weltevrede} P.,  {Lyne} A.~G.,
  {Bassa} C.~G.,  2015, MNRAS, 446, 1380

\bibitem[\protect\citeauthoryear{Radhakrishnan \& Cooke}{Radhakrishnan \&
  Cooke}{1969}]{rc69a}
Radhakrishnan V.,  Cooke D.~J.,  1969, ApL, 3, 225

\bibitem[\protect\citeauthoryear{Rankin}{Rankin}{1983a}]{ran83}
Rankin J.~M.,  1983a, ApJ, 274, 333

\bibitem[\protect\citeauthoryear{Rankin}{Rankin}{1983b}]{ran83a}
Rankin J.~M.,  1983b, ApJ, 274, 359

\bibitem[\protect\citeauthoryear{Rankin}{Rankin}{1986}]{ran86}
Rankin J.~M.,  1986, ApJ, 301, 901

\bibitem[\protect\citeauthoryear{Rankin}{Rankin}{1990}]{ran90}
Rankin J.~M.,  1990, ApJ, 352, 247

\bibitem[\protect\citeauthoryear{Rankin}{Rankin}{1993}]{ran93}
Rankin J.~M.,  1993, ApJ, 405, 285

\bibitem[\protect\citeauthoryear{{Rankin}, {Rodriguez} \& {Wright}}{{Rankin}
  et~al.}{2006}]{rrw06}
{Rankin} J.~M.,  {Rodriguez} C.,    {Wright} G.~A.~E.,  2006, MNRAS, 370, 673

\bibitem[\protect\citeauthoryear{{Rookyard}, {Weltevrede} \&
  {Johnston}}{{Rookyard} et~al.}{2015}]{rwj15}
{Rookyard} S.~C.,  {Weltevrede} P.,    {Johnston} S.,  2015, MNRAS, 446, 3367

\bibitem[\protect\citeauthoryear{Ruderman \& Sutherland}{Ruderman \&
  Sutherland}{1975}]{rs75}
Ruderman M.~A.,  Sutherland P.~G.,  1975, ApJ, 196, 51


\bibitem[\protect\citeauthoryear{{Spitkovsky}}{{Spitkovsky}}{2006}]{spi06}
{Spitkovsky} A.,  2006, ApJL, 648, L51

\bibitem[\protect\citeauthoryear{{Stairs}, {Thorsett} \&
  {Arzoumanian}}{{Stairs} et~al.}{2004}]{sta04}
{Stairs} I.~H.,  {Thorsett} S.~E.,    {Arzoumanian} Z.,  2004, PhRvL, 93, 141101

\bibitem[\protect\citeauthoryear{{Wang}, {Manchester} \& {Johnston}}{{Wang}
  et~al.}{2007}]{wmj07}
{Wang} N.,  {Manchester} R.~N.,    {Johnston} S.,  2007, MNRAS, 377, 1383

\bibitem[\protect\citeauthoryear{{Young}, {Stappers}, {Lyne}, {Weltevrede},
  {Kramer} \& {Cognard}}{{Young} et~al.}{2013}]{ysl+13}
{Young} N.~J.,  {Stappers} B.~W.,  {Lyne} A.~G.,  {Weltevrede} P.,  {Kramer}
  M.,    {Cognard} I.,  2013, MNRAS, 429, 2569

\bibitem[\protect\citeauthoryear{{Yu}, {Manchester}, {Hobbs}, {Johnston},
  {Kaspi}, {Keith}, {Lyne}, {Qiao}, {Ravi}, {Sarkissian}, {Shannon} \&
  {Xu}}{{Yu} et~al.}{2013}]{ymh+13}
{Yu} M.,  {Manchester} R.~N.,  {Hobbs} G.,  {Johnston} S.,  {Kaspi} V.~M.,
  {Keith} M.,  {Lyne} A.~G.,  {Qiao} G.~J.,  {Ravi} V.,  {Sarkissian} J.~M.,
  {Shannon} R.,    {Xu} R.~X.,  2013, MNRAS, 429, 688

\end{thebibliography}
\bibliographystyle{mn2e}

\end{document}